\documentclass[%
reprint,
amsmath,amssymb,
aps,
floatfix,
superscriptaddress
]{revtex4-1}

\usepackage{graphicx}					
\usepackage{dcolumn}					
\usepackage{bm}						
\usepackage{float}
\usepackage{gensymb}
\usepackage{multirow}

\renewcommand{\eqref}[1]{Eq.~(\ref{#1})}
\newcommand{\eqrefs}[2]{Eqs.~(\ref{#1}) and (\ref{#2})}
\newcommand{\eqrefss}[2]{Eqs.~(\ref{#1})--(\ref{#2})}
\newcommand{\eqrefsss}[3]{Eqs.~(\ref{#1}), (\ref{#2}), and (\ref{#3})}
\newcommand{\eqrefssss}[4]{Eqs~(\ref{#1}), (\ref{#2}), (\ref{#3}), and (\ref{#4})}
\newcommand{\Eqref}[1]{Equation~(\ref{#1})}
\newcommand{\Eqrefs}[2]{Equations~(\ref{#1}) and (\ref{#2})}

\newcommand{\Eqrefsss}[3]{Equations~(\ref{#1}), (\ref{#2}), and (\ref{#3})}

\newcommand{\pdf}[2]{\frac{\partial #1}{\partial #2}}

\newcommand{\twovec}[1]{\mathbf{#1}}
\newcommand{\twovecGreek}[1]{\bm{#1}}
\newcommand{\threevec}[1]{\mathbf{#1}}
\newcommand{\threevecGreek}[1]{\bm{#1}}

\renewcommand{\vec}[1]{#1}

\newcommand{\cH}{\vec{c}}

\renewcommand{\i}{\text{in}}
\renewcommand{\o}{\text{out}}
\newcommand{\io}{{\i/\o}}

\newcommand{\B}{\mathrm{f}}
\newcommand{\LB}{g}
\newcommand{\LBd}[2]{\LB_{#2#1}}

\newcommand{\LBp}{\vec{\LB}_P}
\newcommand{\LBps}{\vec{\LB}^*_P}

\newcommand{\LBpi}{\vec{\LB}_P^\i}
\newcommand{\LBpo}{\vec{\LB}_P^\o}

\newcommand{\LBin}{\vec{\LB}^\i}
\newcommand{\LBout}{\vec{\LB}^\o}

\newcommand{\EV}{\bar{g}}

\newcommand{\Nk}{N_K}
\newcommand{\Nt}{N_T}

\newcommand{\Nth}{\frac{\Nt}{2}}

\newcommand{\ep}{E}

\newcommand{\ip}{Q}

\newcommand{\RE}{\text{E}}
\newcommand{\SHE}{\text{sH}}

\newcommand{\mhat}{\hat{\threevec{m}}}

\newcommand{\xhat}{\hat{\threevec{x}}}

\newcommand{\zhat}{\hat{\threevec{z}}}

\newcommand{\khat}{\hat{\mathbf{k}}}
\newcommand{\kp}{\twovec{k}_{||}}
\newcommand{\kvec}{\threevec{k}}

\newcommand{\ch}{c}

\newcommand{\dhat}{\hat{\threevec{d}}}
\newcommand{\fhat}{\hat{\threevec{f}}}
\newcommand{\lhat}{\hat{\bm{\ell}}}

\newcommand{\Gm}{G_{\uparrow \downarrow}}

\newcommand{\ReGm}{\text{Re}[\Gm]}
\newcommand{\ImGm}{\text{Im}[\Gm]}

\newcommand{\ReGS}{G_R}
\newcommand{\ImGS}{G_I}

\newcommand{\total}{\text{tot}}
\newcommand{\interface}{\text{int}}
\newcommand{\magnetization}{\text{mag}}
\newcommand{\lattice}{\text{latt}}
\newcommand{\NM}{\text{NM}}
\newcommand{\HM}{\text{HM}}
\newcommand{\FM}{\text{FM}}

\newcommand{\mf}{\text{mf}}
\newcommand{\sfp}{\text{sf}}

\newcommand{\FSi}{\text{FS}\in\i}

\newcommand{\bulk}{\text{bulk}}
\newcommand{\eq}{\text{eq}}

\newcommand{\lsf}{l_{\text{sf}}}
\newcommand{\tlsf}{t/\lsf}

\newcommand{\IntMinus}[1]{#1(0^-)}
\newcommand{\IntPlus}[1]{#1(0^+)}
\newcommand{\IntPlusMinus}[1]{#1(0^\pm)}
\newcommand{\Int}[1]{#1}

\newcommand{\muM}{\IntMinus{\twovecGreek{\mu}_\perp}}
\newcommand{\muMd}{\IntMinus{\mu_{d}}}
\newcommand{\muMf}{\IntMinus{\mu_{f}}}
\newcommand{\muP}{\IntPlus{\twovecGreek{\mu}_\perp}}

\newcommand{\jM}{\IntMinus{\twovec{j}_\perp}}
\newcommand{\jMd}{\IntMinus{j_{d}}}
\newcommand{\jMf}{\IntMinus{j_{f}}}

\newcommand{\jP}{\IntPlus{\twovec{j}_\perp}}

\newcommand{\jEM}{\IntMinus{\twovec{j}_\perp^\RE}}
\newcommand{\jEMd}{\IntMinus{j^{\RE}_{d}}}
\newcommand{\jEMf}{\IntMinus{j^{\RE}_{f}}}

\newcommand{\jEP}{\IntPlus{\twovec{j}_\perp^\RE}}
\newcommand{\jEPd}{\IntPlus{j^{\RE}_{d}}}

\newcommand{\jEPM}{\IntPlusMinus{\twovec{j}_\perp^\RE}}

\newcommand{\tor}{\Int{\twovecGreek{\tau}}}

\newcommand{\torE}{\Int{\twovecGreek{\tau}^\RE}}

\newcommand{\torEf}{\Int{\tau^{\RE}_f}}

\newcommand{\GR}{\mathbf{G}_R}

\newcommand{\GFM}{\bm{\Gamma}^\FM}
\newcommand{\sigmaMat}{\bm{\sigma}}
\newcommand{\gammaMat}{\bm{\gamma}}

\newcommand{\EF}{\tilde{E}}

\begin{document}

\preprint{APS/123-QED}


\title{Spin Transport at Interfaces with Spin-Orbit Coupling: Phenomenology}


\author{V. P. Amin}
\email{vivek.amin@nist.gov}
\affiliation{
Maryland NanoCenter, University of Maryland, College Park, MD 20742
}
\affiliation{
Center for Nanoscale Science and Technology, National Institute of Standards and Technology, Gaithersburg, Maryland 20899, USA
}
\author{M. D. Stiles}
\affiliation{
Center for Nanoscale Science and Technology, National Institute of Standards and Technology, Gaithersburg, Maryland 20899, USA
}


\date{\today}


\begin{abstract}

This paper presents the boundary conditions needed for drift-diffusion models to treat interfaces with spin-orbit coupling. Using these boundary conditions for heavy metal/ferromagnet bilayers, solutions of the drift-diffusion equations agree with solutions of the spin-dependent Boltzmann equation and allow for a much simpler interpretation of the results.  A key feature of these boundary conditions is their ability to capture the role that in-plane electric fields have on the generation of spin currents that flow perpendicularly to the interface.  The generation of these spin currents is a direct consequence of the effect of interfacial spin-orbit coupling on interfacial scattering. In heavy metal/ferromagnet bilayers, these spin currents provide an important mechanism for the creation of damping-like and field-like torques; they also lead to possible reinterpretations of experiments in which interfacial contributions to spin torques are thought to be suppressed.


\end{abstract}



\maketitle


\section{Introduction}

\begin{figure*}[t!]
	\centering
	\vspace{0pt}	
	\includegraphics[width=1\linewidth,trim={0cm 0cm 0cm 0cm},clip]{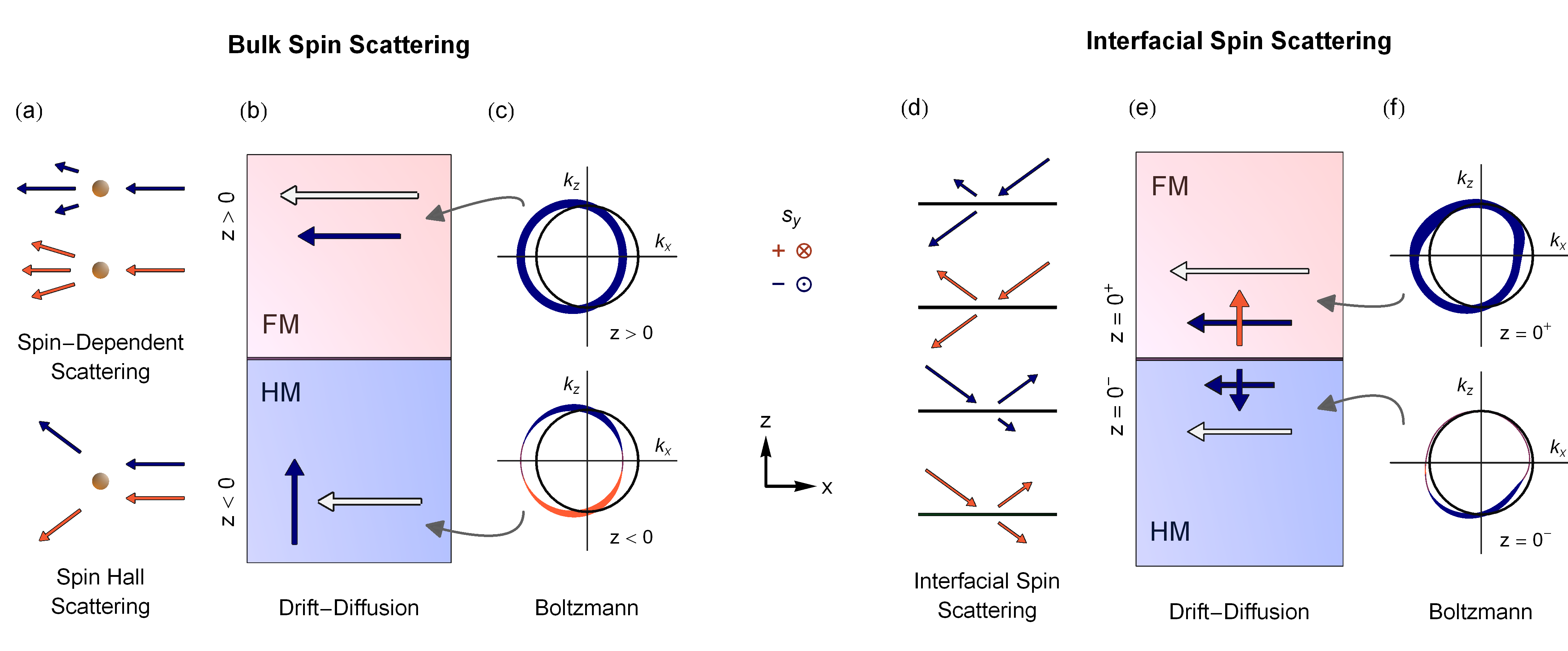}
	\vspace{-3pt}	
	\caption{
	(Color online) Spin and particle currents created by in-plane electric fields in heavy metal/ferromagnet bilayers.  Panel (a) illustrates some scattering processes that give rise to spin currents in the bulk.  Panel (b) depicts these currents within the bilayer as they would be represented in a drift-diffusion approach.  White arrows represent particle currents while the colored arrows represent spin currents.  The direction of these arrows indicates the direction of flow while the color (blue/red) indicates the direction of spin polarization (positive/negative) along the $y$-axis.  Panel (c) shows slices of the non-equilibrium distribution functions that correspond to the currents in panel (b), as described by the Boltzmann equation.  The $\twovec{k}$-space plots each contain the Fermi surface (black circle) and the non-equilibrium distribution functions.  Distortions of the distribution functions outside (inside) the Fermi surface indicate an excess (deficit) of carriers at each $\twovec{k}$-point.  The width of the curves shows the degree of spin polarization and the colors indicate the direction of spin polarization as before.  For example, carriers in the heavy metal moving towards or away the interface have opposite spin polarization; thus the spin Hall current represents a flux of angular momentum (spin current) but no net spin accumulation.  Panel (d) illustrates scattering processes that give rise to spin currents near the interface due to interfacial spin-orbit coupling.  Panel (e) depicts those currents analogously to panel (b), and shows that in-plane charge currents give rise to out-of-plane spin currents that are generated at the interface due to interfacial spin-orbit coupling.  Panel (f) shows the non-equilibrium distribution function corresponding to these currents.  The distribution functions demonstrate that carriers carry a net spin current \emph{and} exhibit a net spin accumulation (unlike the case with the spin Hall effect).  The spin accumulation exerts a torque on the magnetization at the interface via the exchange interaction, while the spin currents exert torques on the neighboring ferromagnet layer via the spin-transfer mechanism.
	}
	\vspace{-3pt}
	\label{fig:motivation}
\end{figure*}

In heavy metal/ferromagnet bilayers, charge currents flowing parallel to the interface can manipulate the magnetization of the ferromagnetic layer \cite{SOTExpAndo, SOTExpMiron, SOTExpGarello, SOTExpFan, SOTExpYu, SOTExpAllen, SOTExpEmori}.  The various mechanisms that drive this process require spin-orbit coupling \cite{SOTTheoryManchon, SOTTheoryManchon2, SOTTheoryMatosAbiague, SOTTheoryHaney}, which couples the spin and orbital moments of carriers.  In addition to this coupling, the orbital moments of carriers are coupled to the crystal lattice via the Coulomb interaction.  Through this extended coupling, carriers receive angular momentum from the atomic lattice and transfer it to the magnetization.  This transfer of angular momentum from carriers to the magnetization is known as a spin-orbit torque \cite{STTTheorySlonczewski, STTTheorySlonczewski2, STTTheoryBerger, STTTheoryRalph, STTTheoryStiles, SOTTheoryManchon, SOTTheoryHaney}.  Spin-orbit torques provide a potentially energy-efficient mechanism to write information to magnetic bits made of heavy metal/ferromagnet bilayers \cite{SOTExpYu}.  

The torques in these bilayers can result from spin-orbit coupling in the bulk and at the interface.  The torques from these two sources have been described in very different ways \cite{SOTExpMiron, SOTExpGarello, SOTExpFan, SOTExpAllen, SOTExpEmori, SOTTheoryHaney}.  The importance of each torque is unclear because of the limited number of models that describe both effects within the same framework \cite{SOTTheoryHaney}.  Since clear phenomenological models can describe the torques created by bulk spin-orbit effects \cite{SOTTheoryHaney}, incorporating interfacial spin-orbit effects into those models will help to properly identify the important mechanisms for spin-orbit torques.  In a companion paper, we introduce a complete phenomenological description of interfacial spin-orbit effects.  In this paper we use the important parts of that description to develop a simple drift-diffusion model for spin-orbit torques in bilayers.

Bulk spin-orbit coupling contributes to spin-orbit torques in heavy metal/ferromagnet bilayers in the following way.  In the heavy metal, bulk spin-orbit coupling causes carriers with opposite spin polarization to scatter in opposite directions.  As a result, charge currents generate spin currents whose polarization and flow directions are orthogonal to each other and to the charge current.  This process, known as the spin Hall effect \cite{SHETheoryDyakonovPerel, SHETheoryHirsch, SHETheoryZhang, SHETheoryMurakami, SHETheorySinova, SHEExpKato, SHEExpWunderlich}, allows for an electric field pointing along the interface to create a spin current that flows across the interface.  Take the interface normal to lie along $\zhat$ and the electric field to point along $\xhat$.  The spin currents that flow along $\zhat$ then generate a flux of angular momentum polarized along the vector $-\xhat \times \zhat$, as illustrated in Fig.~\ref{fig:motivation}(a)-(c).  Upon entering the ferromagnet, this angular momentum is transferred to the magnetization through the spin-transfer mechanism \cite{STTTheorySlonczewski, STTTheorySlonczewski2, STTTheoryBerger, STTTheoryRalph, STTTheoryStiles}.   This process progressively orients the magnetization towards the $-\xhat \times \zhat$ direction, as described by a torque pointing along the direction $\mhat \times [\mhat \times (-\xhat \times \zhat)]$.  Here $\mhat$ denotes the unit vector aligned with the magnetization.  Torques of this form are typically referred to as \emph{damping-like}, since they drive the magnetization towards a particular axis.  In reality, this transfer of angular momentum to the magnetization is not perfect because there is a small component of the spins that rotate when they reflect from the interface, giving rise to torques perpendicular to the damping-like direction \cite{SOTTheoryHaney}.

At the interface between the heavy metal and the ferromagnet, the breaking of inversion symmetry causes an enhanced spin-orbit coupling \cite{REETheoryEdelstein} that leads to a second contribution to spin-orbit torques.  To understand this contribution, note that carriers at the interface develop a net spin accumulation due to a phenomenon known as the Rashba-Edelstein effect \cite{REETheoryEdelstein, REEExpGanichev, REEExpKato, REEExpSilov, REEExpSanchez}.  If this spin accumulation is misaligned with the magnetization at the interface, it exerts a torque on the magnetization via the exchange interaction \cite{SOTTheoryManchon, SOTTheoryManchon2, SOTTheoryMatosAbiague, SOTExpAndo, SOTExpMiron, SOTExpEmori}.  In this geometry, the spin accumulation points along the $-\xhat \times \zhat$ direction; thus the resulting torque on the magnetization points along $\mhat \times (-\xhat \times \zhat)$.  Torques of this form are often referred to as \emph{field-like}, since they force the magnetization to precess around a particular axis.  Typical descriptions of this torque are based on strictly two-dimensional models, which are unrealistic in bilayers because carriers are not actually confined to the interface.  We expect that the spin torques driven by interfacial spin-orbit coupling cannot be quantitatively described by two-dimensional models, since carriers that scatter across the interface behave differently than those that are confined to it.  This suggests that the interfacial contribution to spin-orbit torques requires reexamination using three-dimensional models.

Three-dimensional solutions of the spin-dependent Boltzmann equation show that carriers can exhibit a net spin polarization \emph{and} carry a net spin current near interfaces with spin-orbit coupling.  We illustrate this phenomenon in Fig.~\ref{fig:motivation}(d)-(f).  If the net spin polarization is misaligned with the magnetization, it exerts a torque on the magnetization at the interface.  This captures the spin torque normally associated with the Rashba-Edelstein effect.  However, the spin currents created by interfacial spin-orbit scattering can flow away from the interface, and those that enter the ferromagnet exert additional torques on the magnetization.  These spin currents generate torques via the spin-transfer mechanism, but are driven by interfacial spin-orbit scattering rather than the bulk spin Hall effect.  This mechanism is not usually considered when analyzing spin torques in bilayers, but can contribute significantly to the total spin torque.  It allows for spin torques generated by the interface to have strong damping-like components, which are typically associated with the bulk spin Hall effect.  The spin polarization and flow directions of these spin currents are not required to be orthogonal to each other or to the electric field, unlike the spin currents generated by the spin Hall effect in isotropic bulk systems.

In this three-dimensional picture, one could interpret the net spin polarization as the Rashba-Edelstein effect and the net spin current as an interface-generated spin Hall effect.  First principles calculations support the existence of an interfacial spin Hall effect \cite{SHETheoryWang, SOTTheoryFreimuth2} that could significantly exceed its bulk counterpart \cite{SHETheoryWang}.  Experimental evidence suggests that the spin Hall angle becomes modified near the interface of Bi/Py bilayers, which also alludes to a distinct interfacial contribution to the spin Hall effect \cite{SHEExpHou}.  To assist the interpretation of experiments, the phenomena discussed so far should be incorporated into a simple phenomenological model.  

The drift-diffusion equations are a popular tool used to model transport and analyze experimental results.  They directly relate charge and spin currents to gradients in charge and spin accumulation, but do not describe the momentum-dependence of these quantities.  To treat systems like the bilayers of interest here, the bulk drift-diffusion equations need to be augmented by boundary conditions.  Typically these are taken from magnetoelectronic circuit theory.  However, this approach does not treat interfacial spin-orbit coupling or its consequences.  In the companion paper, we generalize magnetoelectronic circuit theory to include these effects.  Here, we include only the most important changes to magnetoelectronic circuit theory in our boundary conditions when computing spin-orbit torques for a model system.  To test this approach, we compare the results to those found from Boltzmann equation calculations for the same model. 

The Boltzmann equation captures the contributions to transport from carriers at each point in momentum space.  Since spin-orbit scattering is inherently momentum dependent, the Boltzmann equation better describes spin transport in the presence of bulk or interfacial spin-orbit coupling.  For example, the three sources of spin current shown in Fig.~\ref{fig:motivation} can be implicitly captured by the Boltzmann equation \cite{SOTTheoryHaney}.  Solving the Boltzmann equation requires more analytical and computational effort, and is more difficult to directly correlate to experiments.  However, it does provide a good test of the boundary conditions used in the drift-diffusion model, since we can independently calculate the boundary conditions in both models and directly compare the results.

In this paper, we present boundary conditions for drift-diffusion calculations of spin-orbit torques that capture spin-orbit scattering at interfaces.  After introducing these boundary conditions, we use them to solve the drift-diffusion equations for a bilayer system.  This approach gives an analytical model that describes the spin-orbit torques caused by both the spin-Hall and the interfacial Rashba-Edelstein effects.  We then demonstrate that this analytical model predicts spin-orbit torques in quantitative agreement with those found by solving the Boltzmann equation numerically, as long as both methods use the same spin-dependent transmission and reflection coefficients at the interface.

\section{Phenomenology}
\label{sec:Phenom}

In the following we discuss the phenomenology of spin torques in multilayer systems with and without interfacial spin-orbit coupling.  First we consider spin transfer torques in spin valves, and then discuss spin-orbit torques in heavy metal/ferromagnet bilayers.  Throughout this paper we use two coordinate systems: one oriented relative to the interface (to describe electron flow) and the other oriented relative to the magnetization (to describe spin orientation).  In the interface coordinate system, the $x/y$ plane lies along the interface and the $z$ axis points perpendicular to it.  The interface is located at the $z$-axis origin, where $z = 0^-$ and $z = 0^+$ describe the regions just within the non-magnet and ferromagnet respectively.  In the magnetization coordinate system, the direction $\ell$ lies along the magnetization ($\lhat$ = $\mhat$) while the directions $d$ and $f$ are aligned perpendicular to $\lhat$.  Here we choose that the directions $d$ and $f$ point along the vectors $\dhat \propto \mhat \times [\mhat \times (-\xhat \times \zhat)]$ and $\fhat \propto \mhat \times (-\xhat \times \zhat)$ respectively.  As before, we refer to the direction $d$ as \emph{damping-like} and the direction $f$ as \emph{field-like}.  In general, the transverse directions need only span the plane perpendicular to the magnetization.  The transverse directions defined here are merely convenient for describing spin-orbit torques.  


\subsection{Spin Transfer Torque}
\label{sec:PhenomSTT}

We first discuss spin transfer torques in spin valves with no spin-orbit coupling.  Spin valves consist of a non-magnetic metallic spacer sandwiched between two ferromagnetic layers.  The magnetization of one ferromagnetic layer is often fixed via coupling to a neighboring antiferromagnetic layer, while the magnetization of the other layer remains free to change its orientation.  A spin current arises from passing charge current through the fixed layer; this spin current then flows through the non-magnet and transfers angular momentum to the free layer.  

When describing spin accumulations and spin currents in these systems, it is useful to distinguish between those polarized along the magnetization direction and those polarized transversely to it.  At the interface between the non-magnet and the free layer, the spin current polarized along the magnetization direction remains conserved.  However, the spin current with polarization transverse to the magnetization dissipates entirely upon entering the ferromagnet \cite{STTTheoryStiles}.  The interface absorbs part of the transverse spin current, while the remaining portion quickly dissipates within the ferromagnet due to a precession-induced dephasing of spins.  In transition metal ferromagnets and their alloys, this dephasing happens over distances smaller than the spin diffusion length.  Thus we treat the spin accumulation in the ferromagnet as vanishing arbitrarily close to the interface, as is done in magnetoelectronic circuit theory \cite{MCTBrataas, MCTBrataas2}.  The rapid dephasing also allows us to neglect angular momentum transfer (via spin-orbit coupling) to the bulk atomic lattice.  Spin torques can only change the direction of the magnetization, since the magnetization's vector magnitude is considered fixed.  Thus, in the following discussion we only consider spin currents and spin accumulations with polarization transverse to the magnetization.

We refer to the transverse spin current at $z = 0^-$ as $\jM$, where the following expression
\begin{align}
\jM		&=		\GR \muM,	 		\label{eq:jMNoSOC}		
\end{align}
relates this current with the transverse spin accumulation at $z=0^-$  (given by $\muM$).  Here we express the spin accumulation in units of voltage and the spin current in units of number current density \footnote{Multiplying spin currents by $\hbar/2$ redefines them as the angular momentum flux density.  Multiplying by $-g\mu_{\rm B}$,  where $g$ is the electron g-factor, and $\mu_{\rm B}$, the Bohr magneton, converts them into a magnetization flux density.}.    Both the transverse spin current and transverse spin accumulation are two-vectors; in the magnetization reference frame they consist of the damping-like and field-like components of each quantity, i.e. 
\begin{align}
\jM =
\begin{pmatrix}
\jMd		\\
\jMf		\\
\end{pmatrix},
\quad
\muM =
\begin{pmatrix}
\muMd		\\
\muMf		\\
\end{pmatrix}.
\end{align}
The conductance matrix $\GR$ is dependent on the complex-valued \emph{spin mixing conductance} $\Gm$ in the following way:
\begin{align}
\GR =
\begin{pmatrix}
\ReGm		&	-\ImGm		\\
\ImGm		&	\ReGm		\\
\end{pmatrix}.
\end{align}
The spin mixing conductance comes from magnetoelectronic circuit theory and does not depend on the magnetization direction.  

To compute the total spin torque ($\tor^\total$) \footnote{In this paper, torques have units of number current denisty.  To convert them into a form that could be inserted into the Landau-Lifshitz-Gilbert (LLG) equation, one must multiply them by $-g\mu_{\rm B}/t$, where  $t$ is the thickness of the ferromagnetic film.   If the LLG equation is written in terms of the magnetization direction, one should also divide the torques by the saturation magnetization.}, we note that both the interface and the bulk ferromagnet contain magnetization.  Thus, the spin current at $z = 0^-$ equals the flux of angular momentum just outside the ferromagnetic part of the system.  As previously discussed, the interface and the bulk ferromagnet absorb the transverse part of this spin current.  Therefore the total spin torque equals $\jM$, and we only require $\GR$ and $\muM$ to compute it.  

In the ferromagnet, the dephasing processes rapidly destroy the transverse spin accumulations and currents.  This explains why the spin current in \eqref{eq:jMNoSOC} does not depend on $\muP$, as it is negligibly small.  Even though the transverse spin current $\jP$ also dephases, it is useful to identify it as the spin torque on the bulk ferromagnet ($\tor^\FM$).  Thus we may write:
\begin{align}
\tor^\FM		&=		\jP		=		\GFM	 \muM	 		\label{eq:jPNoSOC}
\end{align}
Here the torkance tensor $\GFM$ relates the spin torque in the bulk ferromagnet to the transverse spin accumulation at $z=0^-$.  It has a structure similar to $\GR$, though for now we avoid specifying it.  The spin torque at the interface ($\tor^\interface$) is then the difference between $\tor^\total$ and $\tor^\FM$, which equals the change in transverse spin current from $z = 0^-$ to $z = 0^+$:
\begin{align}
\tor^\interface 		&=		\jM - \jP							\label{eq:IntTorEqualsMagTor}
\end{align}
Using \eqrefs{eq:jMNoSOC}{eq:jPNoSOC} we may then write
\begin{align}
\tor^\interface		&=		\tor^\magnetization	=	(\GR - \GFM) \muM,	 		\label{eq:IntTorNoSOC}
\end{align}
where $\tor^\magnetization$ represents the torque on the magnetization at the interface.  The distinction between $\tor^\interface$ and  $\tor^\magnetization$ is irrelevant in the absence of spin-orbit coupling, since all spin torques in the spin valve are exerted entirely on the magnetization.  However, by introducing interfacial spin-orbit coupling, the magnetization is not the only source of angular momentum that couples to carriers; the lattice provides another source that complicates this analysis and makes this distinction useful.  


\subsection{Spin-orbit Torque}
\label{sec:PhenomSOT}

The need for a fixed ferromagnetic layer is bypassed in heavy metal/ferromagnetic bilayers, where the spin current is generated by the spin Hall effect in the heavy metal.  The spin Hall effect creates spin currents by diverting carriers of charge current with opposite spin in opposite directions.  The spin polarization and flow directions of these spin currents are orthogonal both to each other and to the charge current.  Because carriers flowing in opposite directions carry opposite spin polarization, they contribute to a net spin current but do not exhibit a net spin polarization.  As seen in Fig.~\ref{fig:motivation}(a-c), the electric field that induces the charge current is typically aligned with the interface plane, thus generating a spin current that flows normal to the interface.  The spin torque then arises as it does in spin valves, where the spin current transfers angular momentum to the free layer.

However, spin-orbit scattering near the interface creates spin currents in addition to those caused by the spin Hall effect.  This occurs because individual carriers subject to an in-plane electric field still move in all directions; only their \emph{net} velocity points in-plane.  As a result, carriers scatter off of the interface in a spin-dependent way (due to interfacial spin-orbit coupling) and thus become spin-polarized.  As depicted in Fig.~\ref{fig:motivation}(d-f), the net spin polarization for all carriers does not vanish if the electric field perturbs the occupancy of states differently on each side of the interface.  A difference in the occupancy of states for reflected and transmitted carriers can arise from differing conductivities, degrees of polarization in the ferromagnet, or band structures in each layer.  Unlike the spin Hall effect, carriers subject to interfacial spin-orbit scattering not only carry a net spin current, but also develop a net \emph{spin-polarization}.  This gives carriers two ways to exert a spin torque on the system.

First, we consider the spin currents generated by interfacial spin-orbit scattering.  With the addition of these spin currents, \eqref{eq:jMNoSOC} becomes
\begin{align}
\jM		&=		\GR \muM		+	\jEM				\label{eq:jM}			
\end{align}
The new spin current $\jEM$ may be written as follows
\begin{align}
\jEM		&=		\sigmaMat (\mhat) \EF		,			\label{eq:jEM}			
\end{align}
where the conductivity tensor $\sigmaMat (\mhat)$ vanishes in the absence of interfacial spin-orbit coupling and depends on the magnetization direction $\mhat$.  The scalar $\EF \equiv -E/e$ denotes the in-plane electric field, but is scaled such that the conductivity vector has the same units as the bulk conductivities.  Here we assume that the electric field points along the $x$-axis without loss of generality.  This makes the conductivity tensor a two-vector, although in general the conductivity tensor couples both in-plane electric field components with all spin currents that result from spin-orbit scattering.  We remind the reader that the two-vector $\jEM$ describes spin currents that are polarized transversely to the magnetization and flow perpendicular to the interface plane.  However, these spin currents arise from electric fields that point along the interface plane.  

We note that the conductance tensor $\GR$ is derived in the spirit of magnetoelectronic circuit theory, which means that it does not account for spin-flip scattering at the interface.  Since interfacial spin-orbit coupling leads to spin-flip scattering, the conductance matrix becomes modified as well; however we do not consider such modifications in this paper.  For a simple model of spin-orbit coupling, we show in the companion paper that such modifications only negligibly alter the conductance tensor.  

The transverse spin current at $z = 0^+$ becomes modified by interfacial spin-orbit scattering as well:
\begin{align}
\jP		&=		\GFM \muM		+	\jEP. 			\label{eq:jP}			
\end{align}
The new term is given by
\begin{align}
\jEP		&=		\gammaMat^\FM (\mhat) \EF,		\label{eq:jEP}			
\end{align}
where the \emph{torkivity} tensor $\gammaMat^\FM(\mhat)$ represents the analogue of the conductivity tensor just defined.  As long as the relation $\tor^\FM = \jP$ still holds, the spin current given by \eqref{eq:jEP} now provides an additional contribution to the spin torque on the bulk ferromagnet.

Second, we consider the spin polarization that arises from interfacial spin-orbit scattering.  This spin polarization couples to the magnetization at the interface via the exchange interaction.  This coupling causes carriers to exert an additional spin torque on the magnetization; as a result \eqref{eq:IntTorNoSOC} becomes
\begin{align}
\tor^\magnetization 		&=		(\GR - \GFM) \muM		+	\torE,	 		\label{eq:IntMagTor}
\end{align}
where $\torE$ equals the contribution from interfacial spin-orbit scattering.  We may express this contribution as
\begin{align}
\torE =  \gammaMat^\magnetization(\mhat) \EF		
\label{eq:torE}
\end{align}
where $\gammaMat^\magnetization(\mhat)$ denotes an additional torkivity tensor.  It describes the torque on the magnetization at the interface ($z = 0$), in contrast to the spin current that forms just within the ferromagnet (at $z = 0^+$).  Like all spin currents considered in this paper, these torques have units of number current densities.  

To summarize the results so far, the tensors defined by \eqrefsss{eq:jEM}{eq:jEP}{eq:torE} describe the modifications to spin transport brought upon by interfacial spin-orbit coupling.  The tensors $\sigmaMat (\mhat)$ and $\gammaMat^\FM(\mhat)$ describe the spin currents that arise from spin-orbit scattering near the interface, while $\gammaMat^\magnetization(\mhat)$ describes an additional contribution to the spin torque at the interface.  Each of these tensors may be computed in terms of the spin-dependent reflection and transmission amplitudes at the interface; we provide the necessary expressions in appendix \ref{ap:SimplifiedForm}.  We now discuss how these tensors alter the various spin torques in the bilayer.  

We first remind the reader that $\tor^\interface$ equals the total change in tranverse spin polarization across the interface, while $\tor^\magnetization$ equals the portion of $\tor^\interface$ given to the magnetization.  In the case of the spin valve these torques are identical, as \eqref{eq:IntTorNoSOC} suggests.  However, in heavy metal/ferromagnet bilayers, the interfacial spin-orbit interaction couples carriers to an additional angular momentum bath that is separate from the magnetization.  This suggests that $\tor^\interface$ equals the sum of two torques: one on the magnetization ($\tor^\magnetization$) and the other on the atomic lattice ($\tor^\lattice$).  Thus the interfacial torque now becomes,
\begin{align}
\tor^\interface 		&\equiv	\jM - \jP										\nonumber	\\
 					&=		\tor^\magnetization	+	\tor^\lattice,	 		\label{eq:IntTorBreakdown}
\end{align}
where the lattice torque
\begin{align}
\tor^\lattice 		&=		\tor^\interface - \tor^\magnetization			\nonumber	\\
				&=		\jM - \jP - (\GR - \GFM) \muM - \torE	 		\nonumber	\\
				&=		\jEM - \jEP - \torE							\nonumber	\\
				&=       \left[ \sigmaMat (\mhat) - \gammaMat^\FM (\mhat) -\gammaMat^\magnetization(\mhat) \right] \EF,
				\label{eq:IntLatticeTor}
\end{align}
represents a parasitic contribution to the magnetization torque.  Thus, not only does $\tor^\magnetization$ change in the presence of interfacial spin-orbit coupling (according to \eqref{eq:IntMagTor}), it only partially contributes to the spin torque that carriers exert on the interface ($\tor^\interface$).  


The total spin torque on the magnetization may now be expressed in terms of its interfacial and bulk ferromagnet contributions,
\begin{align}
\tor^\total 		&=		\tor^\magnetization	+	\tor^\FM, 		\nonumber	\\
				&=		\tor^\magnetization	+	\jP,	 			\label{eq:TotTorBreakdownFM}
\end{align}
or by subtracting the lattice torque from the incident flux of angular momentum (i.e. the spin current at $z = 0^-$):
\begin{align}
\tor^\total 		&=		\jM	-	\tor^\lattice.			\label{eq:TotTorBreakdownNM}
\end{align}
\Eqrefs{eq:TotTorBreakdownFM}{eq:TotTorBreakdownNM} represent two separate breakdowns of the total spin-orbit torque, and help to clarify the thickness dependencies of heavy metal/ferromagnet systems.  The spin current in any region vanishes as the layer thickness approaches zero.  Thus, as the ferromagnet thickness vanishes, the total spin-orbit torque approaches the spin torque on the magnetization at the interface ($\tor^{\magnetization}$).  As the heavy metal thickness vanishes, it approaches the opposite of the lattice torque ($-\tor^{\lattice}$) instead.  

\Eqrefsss{eq:jM}{eq:jP}{eq:IntMagTor} capture the phenomenology of interfacial spin orbit coupling and in-plane electric fields.  When used as boundary conditions for the drift-diffusion equations, they allow for quantitative comparison with results from the Boltzmann equation.  



\section{The Drift-Diffusion and Boltzmann Solutions}
\label{sec:BEvsDDE}

We now demonstrate that one may accurately model the interfacial Rashba interaction through the inclusion of the conductivity/torkivity tensors and the mixing conductance.  To study the importance of these parameters we solve the drift-diffusion and Boltzmann equations for a heavy metal/ferromagnet bilayer, using material parameters for a Pt/Co system as found in \cite{SOTTheoryHaney}.  That paper presented a solution to the drift-diffusion equations, but only in the absence of interfacial spin-orbit coupling.  Here, we extend that solution to include interfacial spin-orbit coupling, enabling the calculation of Rashba-based spin-orbit torques.

\subsection{Drift-diffusion solution}
\label{sec:DDESolution}

The drift-diffusion equations directly relate spin and charge accumulations with spin and charge current densities, and do not explicitly treat $\kvec$-dependent scattering.  In the following we describe the three-component spin accumulation and spin current density as $\threevecGreek{\mu}_s$ and $\threevec{j}_s$ respectively.  While the spin current is generally a tensor, here we only consider motion normal to the interface; thus we treat the spin current as a vector in spin space.  The charge accumulation and charge current density are given by $\mu_\ch$ and $j_\ch$.  The latter is a scalar because (as with the spin current) we only consider the out-of-plane current flow.  In this approach (for a spatially-homogenous magnetization $\mhat$) we write the spin current density in the ferromagnet ($z > 0$) as
%
\begin{align}
\threevec{j}_s(z) &= \frac{\sigma^\FM}{e}  P  \mhat \pdf{\mu_\ch(z)}{z} - \frac{\sigma^\FM}{e}\pdf{\threevecGreek{\mu}_s(z)}{z} 		
\label{eq:DDEFMCurrent}
\end{align}
which obeys the following spin continuity equation:
\begin{align}
\frac{1}{eN^\FM_s} \pdf{\threevec{j}_s(z)}{z}	= 	& -\frac{1}{\tau^\FM_{\text{sf}}} \threevecGreek{\mu}_s(z) - \frac{1}{\tau_\text{ex}} \threevecGreek{\mu}_s(z) \times \mhat  					\nonumber	\\
										& -\frac{1}{\tau_\text{dp}} \mhat \times \threevecGreek{\mu}_s(z) \times \mhat.
\label{eq:DDEFMContinuity}
\end{align}
The spin polarization of the current $P$, given by
\begin{align}
P = (\sigma^\FM_\uparrow - \sigma^\FM_\downarrow)/\sigma^\FM
\end{align}
arises because majority and minority carriers have different bulk conductivities.  The right hand side of \eqref{eq:DDEFMContinuity} describes the relaxation due to spin-flip scattering, collective spin precession about the magnetization, and dephasing of the ensemble average of spin, with each mechanism characterized by the time intervals  $\tau^\FM_{\text{sf}}$, $\tau_\text{ex}$, and $\tau_{\text{dp}}$ respectively.  The quantity $N^\FM_s$ is the density of states per unit volume in the ferromagnet.

The corresponding equations for the heavy metal ($z < 0$) contain no magnetization-dependent terms, but include a spin current density source $\threevec{j}_{s}^\text{\SHE} = \sigma_\text{\SHE} \threevec{E} \times \zhat$ to model the spin Hall effect:
\begin{align}
\threevec{j}_s(z) 								&= -\frac{\sigma^\HM}{e} \pdf{\threevecGreek{\mu}_s(z)}{z} + \threevec{j}_{s}^\SHE		\label{eq:DDENMCurrent}			\\
\frac{1}{eN^\HM_s} \pdf{\threevec{j}_s(z)}{z}	&=  -\frac{1}{\tau^\HM_{\text{sf}}} \threevecGreek{\mu}_s(z).							\label{eq:DDENMContinuity}
\end{align}
Here $N^\HM_s$ gives the density of states per unit volume and $\tau^\HM_{\text{sf}}$ equals the spin-flip relaxation time in the heavy metal.  To compute spin-orbit torques, we only need the spin components of all quantities that are transverse to the magnetization.  The drift-diffusion equations that describe these components alone still have the same form as \eqrefs{eq:DDENMCurrent}{eq:DDENMContinuity} in the heavy metal.

According to \eqref{eq:TotTorBreakdownNM}, the total spin-orbit torque may be expressed in terms of $\jM$ and $\tor^\lattice$.  To compute $\tor^\lattice$ we must calculate the conductivity and torkivity matrices given by \eqrefss{eq:apDefSigma1}{eq:apDefGamma}.  To compute $\jM$ we must solve the drift-diffusion equations using the appropriate boundary conditions.  The drift-diffusion equations solved here, as well as the parameters describing the bulk regions, are identical to those used in Ref.~\cite{SOTTheoryHaney}.  However, to capture interfacial spin-orbit effects, we use \eqref{eq:jM} as boundary conditions at $z = 0^-$ instead of magnetoelectronic circuit theory alone.  We also assume that the spin currents vanish at the outer boundaries of both materials.

At $z = 0^+$ we make the approximation that the transverse spin accumulations and currents vanish due to dephasing.  However, our discussion of the total spin-orbit torque in section \ref{sec:PhenomSOT} assumes that the transverse spin current $\jP$ does \emph{not} vanish.  This was necessary so that we analyze the phenomenology of interfacial spin-orbit coupling on both sides of the interface.  Here we only assume that $\jP = \twovec{0}$ in order to simplify the analytical drift-diffusion solution.  We then compute $\tor^\FM$ indirectly by subtracting $\tor^\magnetization$ from $\tor^\total$.  Later we test all of these approximations by comparison to results from the Boltzmann equation.  

In terms of the normal metal thickness $t$, the solution of $\jM$ is given by
\begin{align}
\jM 		&= \mathbf{g}(t) \jEM + \twovec{h}(t) j^\SHE_d													\label{eq:DDESolution}			
\end{align}
Note that the spin Hall current contains no field-like component, so only its damping-like component $j^\SHE_d$ enters this solution.  The matrix $\mathbf{g}(t)$ and the two-vector $\twovec{h}(t)$ have the following structure:
\begin{align}
\mathbf{g}(t) =
\begin{pmatrix}
g_1(t)		&	g_2(t)		\\
g_2(t)		&	g_1(t)		\\
\end{pmatrix}
\quad
\twovec{h}(t) =
\begin{pmatrix}
g_1(t) h_1(t)				\\
g_2(t) h_2(t)				\\
\end{pmatrix}.
\label{eq:DefGh}
\end{align}
The unitless functions $g_1$, $g_2$, $h_1$, and $h_2$ all vanish for zero thickness and converge to finite values for infinite thickness.  To express these functions we define a scaled mixing conductance
\begin{align}
\ReGS 	&\equiv	\ReGm \frac{2 \lsf}{\sigma^{NM}}								\\
\ImGS 	&\equiv	\ImGm \frac{2 \lsf}{\sigma^{NM}} ,	
\end{align}
using the bulk conductivity $\sigma^{\NM}_\bulk$ and the spin diffusion length $\lsf$ of the normal metal.   Then, $G_1$ and $G_2$ are
\begin{align}
g_1(t)	&= \frac{ \tanh^2(\tlsf) - \ReGS \tanh(\tlsf) }{ ( \ReGS - \tanh(\tlsf) )^2 + (\ImGS)^2 }			\label{eq:apG1}		\\
g_2(t)	&= \frac{ \ImGS \tanh(\tlsf) }{ ( \ReGS - \tanh(\tlsf) )^2 + (\ImGS)^2 }	,					\label{eq:apG2}		
\end{align}
which vary monotonically with $\ReGm$ and $\ImGm$ respectively.  The functions $h_1$ and $h_2$
\begin{align}
h_1(t)	&= h_2(t) \frac{ 1 + g_1(t) }{ g_1(t)  }													\label{eq:aph1}		\\
h_2(t)	&= - \frac{ ( 1 - e^{-\tlsf} )^2 }{ 1+ e^{-2\tlsf} }	,										\label{eq:aph2}		
\end{align}
capture extra thickness-dependent terms associated with the spin Hall effect only, as seen in \eqrefs{eq:DDESolution}{eq:DefGh}.  

\begin{figure}[t!]
	\centering
      \vspace{0pt}
	\includegraphics[width=1\linewidth,trim={0.2cm 0cm -0.2cm 0cm},clip]{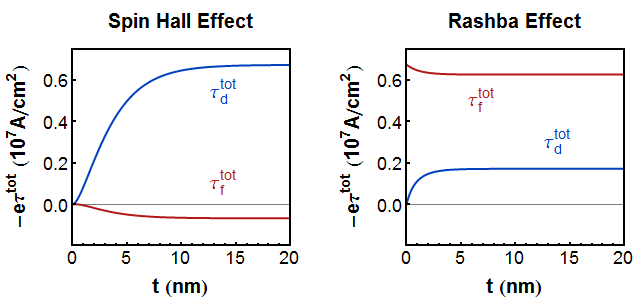}
      \vspace{-5pt}
	\caption{
	(Color online)  Damping-like ($\tau^\total_{d}$) and field-like ($\tau^\total_{f}$) total spin torques plotted versus the heavy metal thickness ($t$).  The spin torques shown originate from either the bulk spin Hall effect or the interfacial Rashba effect.  The ratio of the damping-like and field-like components of the spin Hall torque roughly match the ratio between the real and imaginary parts of the mixing conductance.  Additionally, the spin Hall torque saturates at thicknesses roughly twice that of the Rashba torque.  This thickness-related suppression provides one possible mechanism for Rashba torques to surpass spin Hall torques in thin bilayer systems.  
	}
      \vspace{-2pt}
	\label{fig:DDE}
\end{figure}

According to \eqrefs{eq:TotTorBreakdownNM}{eq:DDESolution}, the total spin torque equals:
\begin{align}
\tor^\total 		&= \mathbf{g}(t) \jEM + \twovec{h}(t) j^\SHE_d - \tor^\lattice.											\label{eq:DDESolutionTotal}			
\end{align}
Without interfacial spin-orbit coupling, the spin current $\jEM$ and the lattice torque $\tor^\lattice$ vanish.  With interfacial spin-orbit coupling, the former contributes to the spin torque thickness dependence while the latter gives the opposite of the zero-thickness intercept.  In particular, $\jEMd$ and $\jEMf$ may be useful fitting parameters for experiments, as they represent the new information required to characterize the thickness dependence of spin-orbit torques.



To compute all boundary parameters we use a scattering potential localized at the interface \cite{SOTTheoryHaney}, based on the Rashba model of spin orbit coupling:
\begin{align}
\label{eq:ScatPot}
V(\threevec{r}) = \frac{\hbar^2 k_F}{m} \delta(z) \big{(} u_0 + u_\text{ex} \threevecGreek{\sigma} \cdot \mhat + u_R \threevecGreek{\sigma} \cdot (\khat \times \zhat) \big{)}
\end{align}
Here $u_0$ represents a spin-independent barrier, $u_\text{ex}$ governs the interfacial exchange interaction, and $u_R$ denotes the Rashba interaction strength.  Plane waves comprise the scattering wavefunctions in both regions.  By deriving reflection and transmission coefficients for majority/minority carriers subject to this interfacial potential, one may compute the conductivity and torkivity tensors using the expressions found in appendices \ref{ap:RTCoeff} and \ref{ap:SimplifiedForm}.  From this we may obtain the parameters $\jEPM$ and $\torE$, which capture the dominant effects of the interfacial spin-orbit interaction.

Fig.~\ref{fig:DDE} shows the total spin torque versus the heavy metal thickness, as caused by the spin Hall and interfacial Rashba effects separately.  As expected, the spin Hall torque shows a mostly damping-like character, while the Rashba torque shows a mostly field-like character.  However, each torque contains both a damping-like and field-like component.  For the spin Hall torque, the ratio between the real and imaginary parts of the spin mixing conductance $\Gm$ roughly determines the ratio between the damping-like and field-like components.  For the Rashba torque, the current sources $\jEPM$ and the lattice torque $\tor^\lattice$ mostly determine this ratio instead.  Interestingly, due to the terms $h_1(t)$ and $h_2(t)$, the spin Hall torque saturates at thicknesses roughly twice that of the Rashba torque.  This thickness-related suppression provides one possible mechanism for Rashba torques to surpass spin Hall torques in thin bilayer systems.  

\begin{figure}[t!]
	\centering
	\includegraphics[width=1\linewidth,trim={0cm 0cm -1cm 0cm},clip]{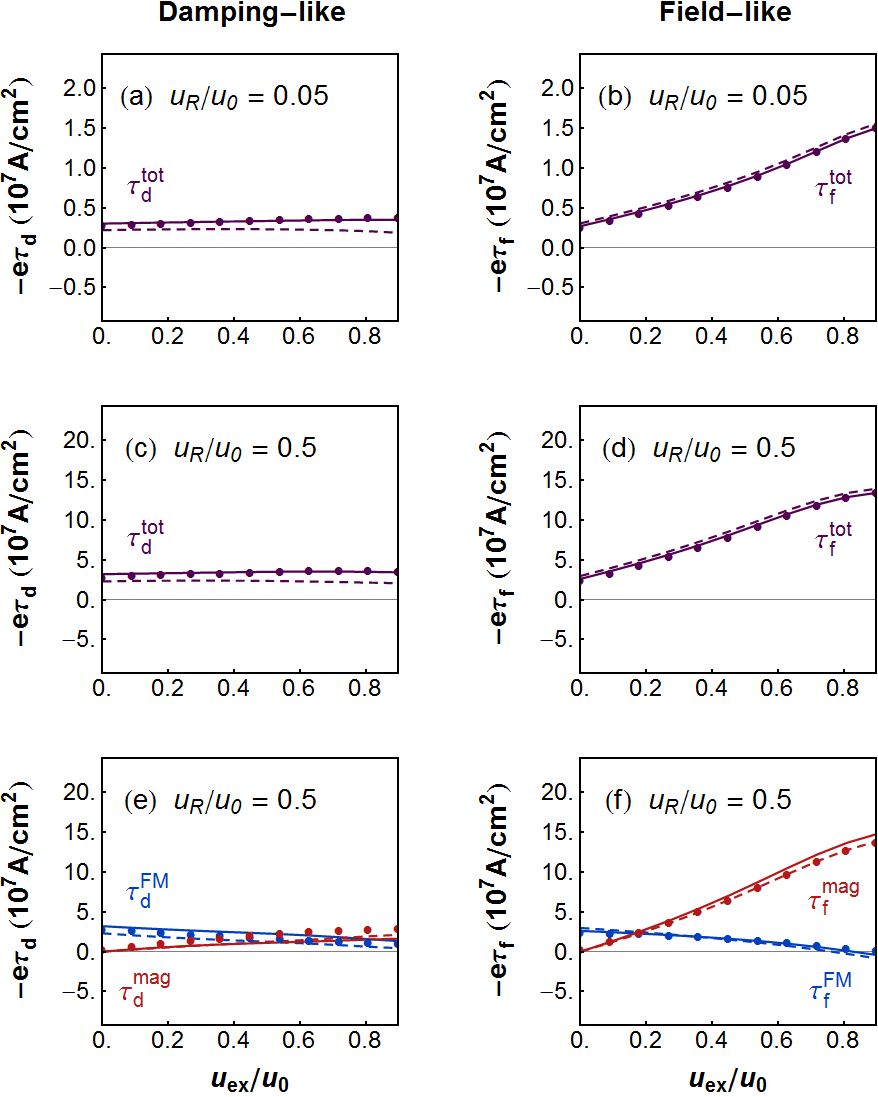}
	\caption{
	(Color online) Spin torques in a Co/Pt bilayer plotted versus the interfacial exchange strength $u_\text{ex}$ (with no spin Hall effect).  The solid curves give the Boltzmann solution, while the dashed curves and the circles give the drift-diffusion/generalized circuit theory solution.  The circles are based on the conductivity and torkivity tensors computed in appendix \ref{ap:SimplifiedForm}, while the dashed curves use more accurately-computed tensors outlined in appendix \ref{ap:BoltzmannSources}.  (a)-(d)  The damping-like ($d$) and field-like ($f$) components of the total spin torque, shown for various $u_\text{R}$.  As $u_\text{ex}$ increases, the total spin torque becomes mostly field-like.  (e)-(f)  Breakdown of the total spin torque into its interfacial (red) and bulk (blue) parts.  For weak $u_\text{ex}$ the bulk spin torque dominates, while for strong $u_\text{ex}$ the interfacial spin torque dominates.  The spin current density $\jEPd$ (which causes a spin torque by flowing into the ferromagnet) significantly contributes to the total damping-like spin torque for weak $u_\text{ex}$.  However, as $u_\text{ex}$ increases the interfacial spin torque must increase as well; eventually its field-like component exceeds all other contributions.  
	}
	\label{fig:Total_Torques}
\end{figure}

\begin{figure*}[t!]
	\centering
      \vspace{0pt}
	\includegraphics[width=1\textwidth,trim={-0.7cm 0cm -0.7cm 0cm},clip]{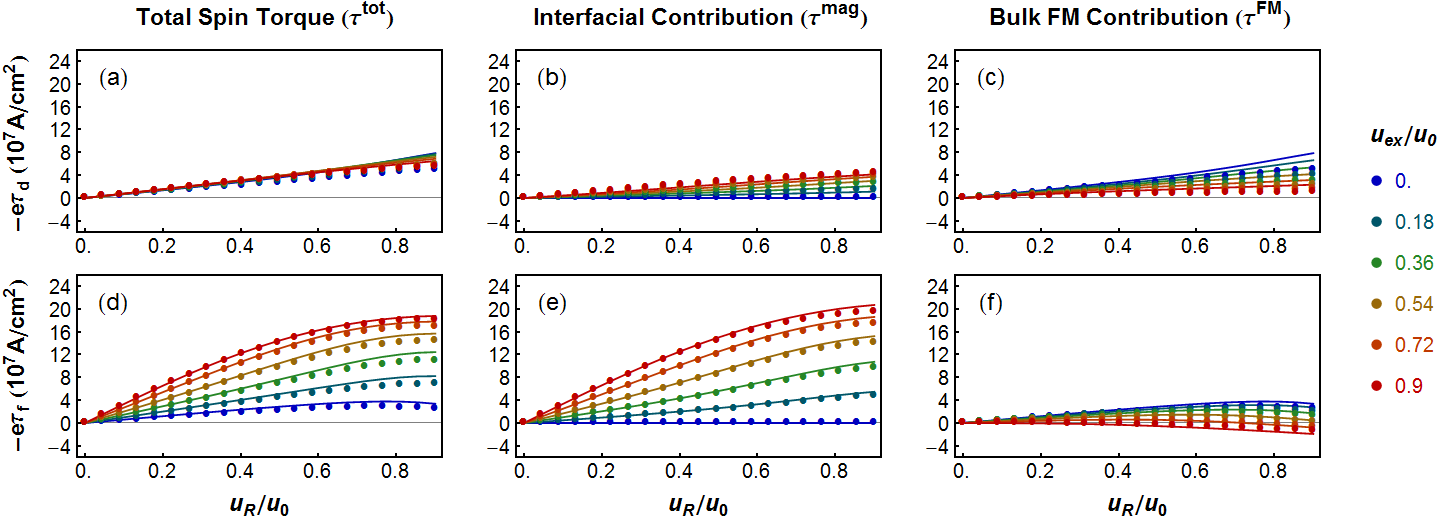}
      \vspace{0pt}
	\caption{
	(Color online) Comparison of the Boltzmann approach (solid curves) and drift-diffusion/generalized circuit theory approach (circles), with the latter using the boundary parameters computed in appendix \ref{ap:SimplifiedForm}.  The panels display spin torques in a Co/Pt bilayer in the absence of the spin Hall effect, plotted versus Rashba parameter ($u_\text{R}$) for various exchange parameter values ($u_\text{ex}$).  Panels (a) and (b) represent the total spin torque, while panels (c)/(d) and (e)/(f) represent the interfacial and bulk contributions respectively.  Both approaches quantitatively agree on the parameterization of the each spin torque, although the drift-diffusion/circuit theory approach slightly underestimates the value of the damping-like spin torque.  The conductivity ($\sigma$) and torkivity ($\gamma$) parameters enable the drift-diffusion equations to describe Rashba spin-orbit torques by capturing the $\kvec$-dependent spin-orbit scattering present in the Boltzmann equation.  Without the inclusion of these parameters, no such drift-diffusion solution exists.
	}
      \vspace{0pt}
	\label{fig:BE_vs_DDE}
\end{figure*}


\subsection{Comparison of the drift-diffusion and Boltzmann approaches}

%
%

To check the validity of the approximations made above we solve the steady-state linearized Boltzmann equation, using the methods described in Refs.~\cite{SOTTheoryHaney},~\cite{Xiao:2007}, and \cite{Stiles:2002a}.  However we do so in the absence of the spin Hall effect, so as to focus on the interfacial Rashba interaction alone.  

Figure~\ref{fig:Total_Torques}(a)-(d) shows the total spin torque versus the interfacial exchange interaction ($u_\text{ex}$) for strong ($u_\text{R}/u_0 = 0.5$) and weak ($u_\text{R}/u_0 = 0.05$) interfacial spin-orbit coupling.  For all cases, the drift-diffusion (circles) and Boltzmann (solid lines) approaches produce quantitatively similar results.  We achieve this agreement by using \eqrefsss{eq:jM}{eq:jP}{eq:IntMagTor} as boundary conditions for the drift-diffusion equations, thus capturing the effects of interfacial spin-orbit scattering.  The conductivity and torkivity tensors that these boundary conditions depend on are derived by approximating the Boltzmann distribution at the interface, as seen in appendix \ref{ap:SimplifiedForm}.  We also present results from an additional drift-diffusion approach (dashed lines) that uses boundary conditions based on a more sophisticated ansatz of the interfacial Boltzmann distribution.  Appendix \ref{ap:BoltzmannSources} outlines the details of this method.  Interestingly, both drift-diffusion approaches agree well with the Boltzmann approach.  For thin layers this agreement may change, since the more sophisticated ansatz of the interfacial Boltzmann distribution takes the outer boundaries into consideration.

Figures~\ref{fig:Total_Torques}(e)-(f) show the interfacial (red) and bulk ferromagnet (blue) contributions to the total spin torque as a function of the strength of the interfacial exchange potential.  The interfacial torque is always field-like, while the bulk ferromagnet torque contains significant damping-like contributions as well.  This occurs because the spin current $\jEPd$ surpasses the spin torque given by $\torEf$ for weak $u_\text{ex}$.  Ordinarily, a damping-like spin torque arises from the spin Hall effect, which does not exist in these results.  However, as $u_\text{ex}$ increases, the damping-like and field-like components of the interfacial spin torque also increase; eventually the field-like component dominates all other spin torque contributions.  This implies that the proximity effect, which could be modeled by $u_\text{ex}$, might change the direction of spin-orbit torques.

Figure~\ref{fig:BE_vs_DDE} compares the Boltzmann approach (solid curves) and the simpler drift-diffusion approach (circles) as a function of the strength of the interfacial spin-orbit coupling. The drift-diffusion solution provides excellent agreement with the Boltzmann solution for all quantities plotted.  This agreement suggests that the conductivity and torkivity tensors would work well as fitting parameters to characterize the impact of interfacial spin-orbit coupling on experimental results.

Figures~\ref{fig:Total_Torques} and \ref{fig:BE_vs_DDE} demonstrate that the boundary conditions given by \eqrefsss{eq:jM}{eq:jP}{eq:IntMagTor} enable the drift-diffusion approach to reproduce results from the Boltzmann approach in the presence of interfacial spin-orbit coupling, despite the fact that the former approach retains no $\kvec$-space information.  The conductivity and torkivity parameters capture interfacial spin-orbit scattering and drive the spin dynamics of spin-diffusive systems; without them the drift-diffusion equations cannot simulate interfacial spin-orbit coupling.  Furthermore, the analytical drift-diffusion solution matches the numerical Boltzmann solution quite well, suggesting that the conductivity and torkivity tensors furnish important parameters when modeling spin-orbit torques.


\section{Outlook}
\label{sec:Outlook}

The conductivity and torkivity tensors capture the physics of interfacial spin-orbit scattering and in-plane electric fields.  In particular, we showed that these tensors strongly influence the potential for a system to produce damping-like and/or field-like torques.  As a result, calculating these tensors for a realistically-modeled system should provide direct insight into its spin transport behavior.  Even so, treating the elements of these tensors as phenomenological parameters should benefit the analysis of a variety of experiments.

Attempts to suppress the Rashba torque in heavy metal/ferromagnet multilayers often involve inserting a metallic spacer between films.  Although this prevents the formation of a single interface with both spin-orbit coupling and an exchange interaction, it creates two interfaces that possess mostly one property or the other.  Fan et al.~\cite{SOTExpFan} measure spin torques in both CoFeB/Pt and CoFeB/Cu/Pt multilayers in order to isolate the spin torque contributions from the heavy metal and from the interface.  To see this, consider the latter system, and note that the Cu spacer prevents the spin polarization at the Cu/Pt interface from directly exerting a torque on the CoFeB layer.  As a result, the spin torque in that system was attributed to the heavy metal, which creates a spin current (via the spin Hall effect) that can pass through the Cu spacer with negligible spin relaxation~\cite{SOTExpFan}.  However, in the interpretation presented here, spin-orbit scattering at the Cu/Pt interface also creates a spin current.  In analogy with the spin Hall effect, this spin current can flow into a neighboring ferromagnetic layer and exert a spin torque.  In general, the polarization direction of this spin current can differ from that generated by the spin Hall effect.  The resulting spin torque is both damping-like and field-like with respect to the field direction $-\threevec{E} \times \zhat$, but is solely damping-like with respect to the polarization direction of the spin current.  Thus, interfaces with spin-orbit coupling could play an active role in generating spin-orbit torques, even when separated from ferromagnetic layers by metallic spacers.

Allen et al.~\cite{SOTExpAllen} measure the Ta thickness dependence for a CoFeB/Ta bilayer and project a non-zero field-like interfacial torque.  The model that they use attributes the thickness dependence only to the spin Hall effect and treats the Rashba torque as an interfacial parameter.  The drift-diffusion solution presented in section \ref{sec:DDESolution} provides a generalization of this analysis in the presence of Rashba spin orbit coupling and a possible explanation of the zero-thickness intercept.

Finally, Garello et al.~\cite{SOTExpGarello} measure strongly anisotropic damping-like and field-like torques that depend heavily on growth techniques and material composition.  We note that the scattering amplitudes considered depend on magnetization direction and interfacial disorder, which lead to such anisotropy within the boundary parameters.  Further work is required to characterize this anisotropy.

We expect that the most useful approach for interpreting experiments as above is to treat the new transport parameters as fitting parameters.  In the future, this approach can be checked by calculating the parameters from first principles \cite{IntResStiles,Xia:2002} as has been done for magnetoelectronic circuit theory.  In the companion paper we generalize the expressions given by \eqrefss{eq:apDefSigma1}{eq:apDefGamma} for the case of realistic electronic structures.  Such calculations would provide a useful bridge between direct first-principles calculations of spin torques \cite{SOTTTheoryHaney2,SOTTheoryFreimuth,SOTTheoryFreimuth2,SOTTheoryFreimuth3} and drift-diffusion calculations done to analyze experiments.


To conclude, we present boundary conditions that capture the phenomenology of interfacial spin-orbit scattering when driven by in-plane electric fields, which was previously inaccessible to the drift-diffusion equations.  Using these boundary conditions we solve the drift-diffusion equations for a bilayer system, and obtain an analytical model of spin-orbit torques caused by both the spin Hall and Rashba-Edelstein effects.  We then compare the spin-orbit torques predicted by this drift-diffusion approach with those obtained by solving the spin-dependent Boltzmann equation.  We find quantitative agreement between both approaches over a large parameter space, which includes both strong and weak interfacial spin-orbit coupling.  Most importantly, we find that the spin currents created by interfacial spin-orbit scattering must be considered to achieve agreement between these approaches.  Finally, we discuss the interpretation of current experiments, and describe in particular how an interface can exert a spin torque on a nearby ferromagnetic layer without being directly connected to it.


\begin{acknowledgments}
The authors thank Kyoung-Whan Kim, Paul Haney, Guru Khalsa, Kyung-Jin Lee, and Hyun-Woo Lee for useful conversations and Robert McMichael and Thomas Silva for critical readings of the manuscript.  VA acknowledges support under the Cooperative Research Agreement between the University of Maryland and the National Institute of Standards and Technology, Center for Nanoscale Science and Technology, Grant No. 70NANB10H193, through the University of Maryland.
\end{acknowledgments}


\appendix

\section{Derivation of reflection and transmission matrices}
\label{ap:RTCoeff}
%
Interfacial spin-orbit coupling causes both momentum and spin-dependent scattering at interfaces.  If the incident distribution of carriers depends on momentum and/or spin, outgoing carriers may become spin-polarized via interfacial spin-orbit scattering.  This gives rise to non-vanishing spin accumulations, spin currents, and spin torques, which are related by \eqrefsss{eq:jM}{eq:jP}{eq:IntMagTor}.  In order to derive the tensors introduced in those expressions, we must first describe how an ensemble of spins scatters at an interface.

One may relate the spinors describing carriers incident ($\chi$) and scattered away from ($\xi$) an interface by the following relation,
\begin{align}
\xi = r \chi
\end{align}
subject to the $2 \times 2$ reflection matrix for majority/minority spin states:
\begin{align}
\label{eq:RTamplitudes}
r =
\begin{pmatrix}
r^\uparrow					&	r^{\uparrow\downarrow}	\\
r^{\downarrow\uparrow}		&	r^\downarrow				\\
\end{pmatrix}.
\end{align}
For now we omit the transmission of carriers from the opposite side of the interface.  Given the density matrix associated with an ensemble of incident carriers,
\begin{align}
\rho^\i = \sum_n p_n \chi_n \chi_n^\dagger,
\end{align}
where $p_n$ denotes the probability of occupying the spin state $\chi_n$, the density matrix for outgoing carriers becomes:
\begin{align}
\rho^\o = \sum_n p_n \xi_n \xi_n^\dagger = r \rho^\i r^\dagger.
\end{align}
Since density matrices are Hermitian one may expand them as follows
\begin{align}
\rho = \LB_\ch \sigma_\ch + \LB_s \sigma_s.
\end{align}
where $\sigma_s$ denote the Pauli matrices ($s \in [x,y,z]$) and $\sigma_\ch = I_{2 \times 2}$.  One may show that the coefficients $\LB_s$ equal the ensemble average of spin in direction $s$, while $\LB_c$ gives the total probability of occupation.  The outgoing density matrix then becomes
\begin{align}
\rho^\o = \LB^\o_\alpha \sigma_\alpha = \LB^\i_\beta r \sigma_\beta r^\dagger,
\end{align}
where the $\alpha \in [x,y,z,\ch]$.  By obtaining the matrix $R_{\alpha\beta}$ such that
\begin{align}
\LB^\o_\alpha = R_{\alpha\beta} \LB^\i_\beta,
\end{align}
one characterizes the scattering of an ensemble of spins in the semiclassical limit (in terms of scattering amplitudes for minority/majority carriers alone).  Using the identity
\begin{align}
\mathrm{tr}[\sigma_\alpha \sigma_\beta] = 2 \delta_{\alpha\beta} I_{2 \times 2}
\end{align}
one may show that the $R_{\alpha\beta}$ becomes
\begin{align}
R_{\alpha \beta} &= \frac{1}{2} \mathrm{tr} [r^\dagger \sigma_\alpha r \sigma_\beta].
\end{align}
The matrix that describes transmission ($T_{\alpha\beta}$) may be obtained in a similar fashion.  If the scattering amplitudes couple additional channels (such as in-plane momentum or orbital quantum number) we may write
\begin{equation}
S^{XY}_{mn,\alpha \beta} =
\begin{cases} 
R^{X}_{mn,\alpha \beta}		\quad \quad	\text{for }	X = Y		\\
T^{XY}_{mn,\alpha \beta}		\quad \quad	\text{for }	X \neq Y		.
\end{cases}
\label{eq:RTBig}
\end{equation}
for
\begin{align}
R^{X}_{mn,\alpha \beta} &= \frac{1}{2} \mathrm{tr} \big{[} (r^X_{mn})^\dagger \sigma_\alpha r^X_{mn} \sigma_\beta \big{]}	\\
T^{XY}_{mn,\alpha \beta} &= \frac{1}{2} \mathrm{tr} \big{[} (t^{XY}_{mn})^\dagger \sigma_\alpha t^{XY}_{mn} \sigma_\beta \big{]},
\end{align}
instead, where $m$/$n$ label the additional channels and $X,Y \in [+,-]$ label the sides of the interface.  These scattering matrices comprise boundary conditions suitable for semiclassical models such as the Boltzmann equation.

\section{Derivation of the conductivity and torkivity parameters}
\label{ap:SimplifiedForm}
%
%
We now derive expressions for the conductivity and torkivity parameters introduced in \eqrefsss{eq:jEM}{eq:jEP}{eq:torE}, which can be expressed in terms of scattering amplitudes.  To do so we use the non-equilibrium distribution function $\LB_\alpha(\threevec{r},\threevec{k})$, which describes a perturbation of the equilibrium Fermi-Dirac distribution $\B_{\eq}(\varepsilon_\mathbf{k})$ and depends on the position, momentum, and spin of carrier wavepackets:
\begin{align}
\B_\alpha (\threevec{r},\threevec{k}) = \B_{\eq}(\varepsilon_\mathbf{k}) \delta_{\alpha\ch} + \pdf{\B_{\eq}}{\varepsilon_\mathbf{k}} \LB_\alpha(\threevec{r},\threevec{k}).
\label{eq:apBoltDist}
\end{align}
The representation that we use includes four distributions ($\alpha \in [d,f,\ell,\ch]$), where the first three refer to spins polarized along each axes and the last one refers to the total population.  Note that in the following we refer to $\alpha$ as a spin/charge index.  We approximate the portion of $\LB_\alpha(\threevec{r},\threevec{k})$ incident to the interface in the non-magnet as follows:
\begin{align}
\LB^{\i}_\alpha(0^-,\kp) 	=	&~	e^2 \EF \tau^\NM v_x(\kp) \delta_{\alpha \ch}.					\label{eq:apIncDist1}
\end{align}
Note that $\kp$ denotes the in-plane momentum vector, $\tau^\NM$ gives the the momentum relaxation time in the non-magnet, and $\EF = -E/e$ equals the scaled in-plane electric field.  Without loss of generality we assume that the electric field points along the $x$-axis. This distribution function weights the occupancy of carriers such that those moving in the direction opposite to the electric field outnumber those moving in the same direction.  Therefore \eqref{eq:apIncDist1} captures the charge current that arises from an electric field.  In a ferromagnet, an electric field creates both a charge and spin current, captured by the following distribution function instead:
\begin{align}
\LB^{\i}_\alpha(0^+,\kp) 	=	&~ e^2 \EF \tau^\FM v_x(\kp)									\nonumber			\\
							&	\times \Big{(} \delta_{\alpha \ch} - P \delta_{\alpha \sigma} \hat{m}_\sigma \Big{)}.			\label{eq:apIncDist2}
\end{align}
Here the index $\sigma \in [d,f,\ell]$ runs over only the spin components.  The quantity $\tau^\FM$ gives the momentum relaxation time in the ferromagnet, while 
\begin{align}
P = \big{(} \sigma^\FM_\uparrow - \sigma^\FM_\downarrow \big{)} / \sigma^\FM,
\end{align}
equals the polarization in the ferromagnet, given in terms of the bulk conductivities for majority and minority carriers.

\Eqrefs{eq:apIncDist1}{eq:apIncDist2} give the anisotropic contributions to the non-equilibrium distribution function caused by an in-plane electric field.  They are derived from the particular solution of the Boltzmann equation in the relaxation time approximation.  Here, for numerical simplicity, we assume the same spherical Fermi surface describes both regions \cite{SFSATheoryZhang} and both spins in the ferromagnet.  The spin-dependent conductivity in the ferromagnetic material is captured by different scattering times for majority and minority carriers.  In appendix \ref{ap:BoltzmannSources}, we generalize the expressions presented in this appendix to describe non-trivial electronic structures.  

The momentum relaxation times used in \eqrefs{eq:apIncDist1}{eq:apIncDist2} are renormalized by bulk spin-flip scattering in the non-magnet and account for differing majority and minority relaxation times in the ferromagnet:
\begin{align}
\label{eq:apTransTimes}
(\tau^\NM)^{-1} 	&= (\tau^\NM_{\mf})^{-1} + (\tau^\NM_{\sfp})^{-1}							\\
(\tau^\FM)^{-1} 	&= 2 (\tau^{\FM\uparrow}_{\mf} + \tau^{\FM\uparrow}_{\mf})^{-1}
\end{align}
For the non-magnet, $\tau^\NM_{\mf}$ denotes the mean free scattering time while $\tau^\NM_{\sfp}$ denotes the spin-flip scattering time.  For the ferromagnet, $\tau^{\FM\uparrow}_{\mf}$ and $\tau^{\FM\downarrow}_{\mf}$ denote the mean free scattering times for majority and minority carriers respectively.  We may better approximate \eqrefs{eq:apIncDist1}{eq:apIncDist2} by forcing the distribution function to obey outer boundary conditions as well.  In appendix \ref{ap:BoltzmannSources} we present a more sophisticated approximation for \eqrefs{eq:apIncDist1}{eq:apIncDist2} that accomplishes this by incorporating solutions to the homogeneous Boltzmann equation.

The outgoing distribution at $z = 0^-$ is specified by the incoming distributions of both sides and interfacial scattering coefficients:
\begin{align}
\LB^{\o}_{\alpha}(0^-,\kp)	&=		R^-_{\alpha \beta}(\kp) \LB^{\i}_{\beta}(0^-,\kp)						\nonumber			\\
								&+		T^{-+}_{\alpha \beta}(\kp) \LB^{\i}_{\beta}(0^+,\kp).			\label{eq:apLBOut1}
\end{align}
Note that the scattering coefficients depend on magnetization in general.  Likewise, the outgoing distribution at $z = 0^+$ is expressed as follows:
\begin{align}
\LB^{\o}_{\alpha}(0^+,\kp)	&=		R^+_{\alpha \beta}(\kp) \LB^{\i}_{\beta}(0^+,\kp)						\nonumber			\\
								&+		T^{+-}_{\alpha \beta}(\kp) \LB^{\i}_{\beta}(0^-,\kp).				\label{eq:apLBOut2}
\end{align}
To calculate non-equilibrium quantities on either side of the interface, we must compute integrals of the distribution function over the Fermi surface (FS).  In terms of the incoming and outgoing distribution functions, the spin current densities $j^\RE_{\sigma}$ ($\sigma \in [d,f,\ell]$) on each side of the interface are
\begin{align}
j^\RE_{\sigma}(0^-) 	=	- \frac{c}{e v_F} \int_{\FSi} d^2k v_{z}	\Big{[}	&		\big{(} R^-_{\sigma \beta} - \delta_{\sigma\beta} \big{)} \LB^{\i}_{\beta}(0^-) 			\nonumber				\\
																	&+		T^{-+}_{\sigma \beta} \LB^{\i}_{\beta}(0^+)	\Big{]}										\label{eq:apDefCurrent1}	\\
j^\RE_{\sigma}(0^+) 	=	- \frac{c}{e v_F} \int_{\FSi} d^2k v_{z}	\Big{[}	&		\big{(} R^+_{\sigma\beta} - \delta_{\sigma \beta} \big{)} \LB^{\i}_{\beta}(0^+) 		\nonumber				\\
																	&+		T^{+-}_{\sigma \beta} \LB^{\i}_{\beta}(0^-)	\Big{]},										\label{eq:apDefCurrent2}	
\end{align}
where the constant $c$ is given by:
\begin{align}
c &= -\frac{e}{\hbar} \frac{1}{(2\pi)^3}.			
\end{align}
Here we write $j^\RE_{\sigma}$ in units of number current density.  The phase-coherent spin densities on each side of the interface are
\begin{align}
\langle s^\RE_\sigma (0^-) \rangle		&=  - \frac{c}{e v_F} \int_{\FSi} d^2k T^{+-}_{\sigma \alpha} \LB^{\i}_{\alpha}(0^-)		\label{eq:apDefTorque1}	\\
\langle s^\RE_\sigma (0^+) \rangle		&=  - \frac{c}{e v_F} \int_{\FSi} d^2k T^{-+}_{\sigma \alpha} \LB^{\i}_{\alpha}(0^+).		\label{eq:apDefTorque2}
\end{align}
We write these spin densities in units of number density.  The total spin density $\langle \threevec{s}^\RE \rangle$ equals the sum of the contributions from both sides:
\begin{align}
\langle \threevec{s}^\RE \rangle = \langle \threevec{s}^\RE(0^-) \rangle + \langle \threevec{s}^\RE(0^+) \rangle. 	\label{eq:apDefTorque4}	
\end{align}
The spin torque on the magnetization at the interface is given by 
\begin{align}
\tau^\RE = - \frac{ J_\text{ex} }{ \hbar } \langle \threevec{s}^\RE \rangle \times \mhat,		\label{eq:apDefTorque3}	\\[-0.1cm]	\nonumber
\end{align}
where $J_\text{ex}$ equals the exchange energy at the interface.  For the scattering potential given by \eqref{eq:ScatPot}, $J_\text{ex}$ becomes
\begin{align}
J_\text{ex} = - \frac{ \hbar k_F u_\text{ex} }{ m }.
\end{align}
It is convenient to write the spin current density and spin torque in terms of the conductivity $\sigma_{\sigma}$ and torkivity $\gamma_\sigma$ parameters:
\begin{align}
j^E_{\sigma}(0^-) &= \sigma_{\sigma} \EF 						\label{eq:apJQ}	\\
j^E_{\sigma}(0^+) &= \gamma^\FM_{\sigma} \EF					\label{eq:apJQ}	\\
\tau^E_{\sigma} &= \gamma^\magnetization_{\sigma} \EF	. 		\label{eq:TauapQ}
\end{align}
Using \eqrefssss{eq:apIncDist1}{eq:apIncDist2}{eq:apLBOut1}{eq:apLBOut2} to evaluate \eqrefss{eq:apDefCurrent1}{eq:apDefTorque4}, one may express these tensors in terms of the magnetization-dependent scattering coefficients:
\begin{widetext}
\begin{align}
\sigma_{\sigma}(0^-) 	&=	-\frac{ec}{v_F} \int_{\FSi}	d^2k v_{z} v_{x}	
														\Big{[}		
																	\tau^\NM \big{(} R^-_{\sigma \ch} - \delta_{\sigma\ch} \big{)} 						
																+	\tau^\FM \big{(} T^{-+}_{\sigma \ch} 
																-	P T^{-+}_{\sigma \sigma'} \hat{m}_{\sigma'} \big{)}					
														\Big{]},																										\label{eq:apDefSigma1}	\\
\gamma^\FM_{\sigma}	&=	-\frac{ec}{v_F} \int_{\FSi}	d^2k v_{z} v_{x}	
														\Big{[}		
																	\tau^\NM T^{+-}_{\sigma \ch}	
																+	\tau^\FM \big{(} R^+_{\sigma \ch} - \delta_{\sigma\ch}
																-	P ( R^+_{\sigma \sigma'} - \delta_{\sigma \sigma'} ) \hat{m}_{\sigma'} \big{)} 												
														\Big{]},																										\label{eq:apDefSigma2}	\\
\gamma^\magnetization_{\sigma} 		&= 	- u_\text{ex} ec \int_{\FSi} d^2k v_{x} \epsilon_{\sigma\sigma'\sigma''} \hat{m}_{\sigma'} 													
														\Big{[}		
																	\tau^\NM T^{+-}_{\sigma''\ch}																			
																+	\tau^\FM \big{(} T^{-+}_{\sigma''\ch} - P T^{-+}_{\sigma''\sigma} \hat{m}_\sigma \big{)}	
														\Big{]}.																										\label{eq:apDefGamma}			
\end{align}
\end{widetext}
For $\sigma \in [d,f]$ we produce the tensors introduced in section \ref{sec:PhenomSOT}.  In the same spirit as magnetoelectronic circuit theory, these tensors represent moments of the scattering coefficients weighted by velocities.  Note that for $P = 0$ the tensors describing spin currents do not vanish, so long as the momentum relaxation times of each region differ and carriers are subject to interfacial spin-orbit scattering.  This suggests that non-magnetic interfaces with spin-orbit coupling still behave as sources of spin current.

\section{The discretized Boltzmann equation}
\label{ap:Boltzmann}
%
The spin-dependent Boltzmann equation is given by
\begin{align}
\frac{\partial \B_\alpha}{\partial t} 
+ \pdf{\threevec{r}}{t} \pdf{\B_\alpha}{\threevec{r}} 
+ \pdf{\threevec{k}}{t} \pdf{\B_\alpha}{\threevec{k}}
+ \gamma \epsilon_{\alpha \beta \gamma} H^{\text{ex}}_\beta \B_\gamma
=
\pdf{\B_\alpha}{t}_{coll}.
\end{align}
where Greek letters label spin/charge indices ($\alpha, \beta \in [d,f,\ell,\ch])$ and are implicitly summed over unless otherwise stated.  The fourth term, however, describes spin precession in a ferromagnet and excludes the charge distribution from the implicit sums.  One may use the semiclassical equations of motion to determine the following time derivatives
\begin{align}
\pdf{\threevec{r}}{t} &= \threevec{v}(\threevec{k}) \\
\pdf{\threevec{k}}{t} &= -e\threevec{E}
\end{align}
where $\threevec{v}$ denotes the electron velocity and $\threevec{E}$ equals the electric field.  In the limit that the distribution functions are small perturbations of the Fermi function, i.e.
\begin{align}
\B_\alpha (\threevec{k}) \rightarrow \B_{eq}(\varepsilon_\threevec{k}) \delta_{\alpha\ch} + \pdf{\B_{eq}}{\varepsilon_\threevec{k}} \LB_\alpha(\threevec{r},\threevec{k}) 
\end{align}
we obtain the linearized Boltzmann equation (in steady-state)
\begin{align}
\label{eq:LBEap}
&v_z(\threevec{k}) \pdf{\LB_\alpha(\threevec{k})}{z}
- e \threevec{E} \cdot \threevec{v}_x(\threevec{k}) \delta_{\alpha\ch}
+ \gamma \epsilon_{\alpha \beta \gamma} H^{\text{ex}}_\beta \LB_\gamma(\threevec{k}) 	\nonumber	\\
&= 
-R_{\alpha\alpha'}(\threevec{k}) \LB_{\alpha'}(\threevec{k})
+ \int_{FS} d\threevec{k}' P_{\alpha\alpha'}(\threevec{k}, \threevec{k}') \LB_{\alpha'}(\threevec{k}')
\end{align}
assuming that any position-dependence is restricted to the $z$ axis.  The latter assumption applies to systems with translational-invariance in the $x/y$ plane.  Note that all $\threevec{k}$ vectors are limited to the Fermi surface.  

We now treat the Fermi surface as a mesh of $\Nk$ discrete vectors, labeled by some index $i$.  Using the following prescription
\begin{align}
\vec{k} &\rightarrow i \\
\LB_\alpha(\vec{k}) &\rightarrow \LBd{\alpha}{i} \\
R_{\alpha\alpha'}(\vec{k}_i) &\rightarrow R_{i,\alpha\alpha'} \\
P_{\alpha\alpha'}(\vec{k}_i, \vec{k}_j) &\rightarrow P_{ij,\alpha\alpha'} \\
\int_{FS} h_\alpha(\vec{k}) d\vec{k} &= \sum_{i=1}^{N_k} w_i h^\alpha_i
\end{align}
we may write \eqref{eq:LBEap} as
%
%
\begin{align}
\label{eq:DBE1}
\pdf{\LBd{\alpha}{i}}{z} + \sum_j B_{ij,\alpha\alpha'} \LBd{\alpha'}{j}
= 
e E \delta_{\alpha\ch} \frac{v_{x,i}}{v_{z,i}}  
\end{align}
where
\begin{align}
B_{ij,\alpha\alpha'}
\equiv
\frac{1}{v_{z,i}} 
\bigg{[}
&\gamma \epsilon_{\alpha \beta \alpha'} H^{ex}_\beta \delta_{ij} \nonumber \\
&+ R_{i,\alpha\alpha'} \delta_{ij}
- w_j P_{ij,\alpha\alpha'}
\bigg{]}.
\end{align}
Here we assume that $\threevec{E} = E\xhat$.  Note that $w_i$, which transforms any sum into a Fermi surface integral, depends on the mesh choice.  Combining the indices $i$ and $\alpha$ into a single index, we may write \eqref{eq:DBE1} in vector form
\begin{align}
\label{eq:DBE2}
\pdf{\vec{\LB}}{z} + B \vec{\LB} = E \vec{v_x^*}
\end{align}
using the definition
\begin{align}
[\vec{v_x^*}]^i_{\alpha} \equiv e \frac{v_{x,i}}{v_{z,i}} \delta_{\alpha\ch}
\end{align}
where both $\vec{\LB}$ and $\vec{v_x^*}$ contain $\Nt \equiv 4 \times \Nk$ elements, making $B$ an $\Nt \times \Nt$ matrix.  The full solution then becomes
\begin{align}
\label{eq:LBdSol}
\vec{\LB} = \LB_{P} + \LB_H
\end{align}
which satisfy
\begin{align}
\label{eq:LBpSol}
B \LB_{P} &= E \vec{v_x^*}
\end{align}
and
\begin{align}
\label{eq:LBhSol}
\LB_H =\sum_n c_n e^{\lambda_n z} \vec{\EV}_n
\end{align}
where $\lambda_n$ and $\vec{\EV}_n$ are respectively the $\Nt$ eigenvalues and eigenvectors of the $B$ matrix.  The particular solution $\LB_{P}$ describes the direct response to an external electric field, whereas $\LB_H$ represents a linear combination of the $\Nt$ solutions to the homogenous Boltzmann equation.  Both boundary conditions and the external electric field determine the coefficients $c_n$ \footnote{
The eigenvectors $\vec{\EV}_n$ come in pairs with eigenvalues of same magnitude but opposite sign, except in the case of a vanishing eigenvalue.  Those solutions are paired instead with
\begin{align}
\vec{\EV}_{0'} = z \vec{\EV}_0 - \vec{\EV}_{0''}
\end{align}
where $\vec{\EV}_0$ denotes any solution with zero eigenvalue, $\vec{\EV}_{0'}$ gives its pair solution, and $B \vec{\EV}_{0''} = \vec{\EV}_0$.  Vanishing eigenvalues occur, for instance, within the relaxation time approximation because the matrix $B$ is singular.}.  Equation~(\ref{eq:LBhSol}) implies that all solutions to the homogenous equation vary exponentially over position, but may possess some complicated spin-dependent distribution over $\kvec$-space.  

We may also write \eqref{eq:LBhSol} as
\begin{align}
\label{eq:LBhSol1}
\LB_H = Z \Lambda(z) \cH
\end{align}
where $Z$ is an $\Nt \times \Nt$ matrix defined by the column vectors $\vec{\EV}_n$,
\begin{align}
\label{ZDef}
Z =
\begin{pmatrix}
\vec{\EV}_{1} 	& 	\vec{\EV}_{2}	&	\cdots	&	\vec{\EV}_{\Nt}
\end{pmatrix}
\end{align}
$\Lambda(z)$ describes the position dependence,
\begin{align}
\Lambda(z) =
\begin{pmatrix}
e^{\lambda_1 z} 	& 	0				&	\cdots	&	0					\\
0		 		& 	e^{\lambda_2 z}	&	\cdots	&	0					\\
\vdots		 	& 	\vdots			&	\ddots	&	\vdots 				\\
0			 	& 	0				&	\cdots	&	e^{\lambda_{\Nt} z}
\end{pmatrix}
\end{align}
and $\cH$ is a vector containing the coefficients of expansion for the homogeneous solutions:
\begin{align}
\cH =
\begin{pmatrix}
\label{cDef}
c_{1} 	& 	c_{2}	&	\cdots	&	c_{\Nt}
\end{pmatrix}
\end{align}
At $z=0$, $\Lambda$ yields the identity matrix and \eqref{eq:LBdSol} becomes
\begin{align}
\label{eq:LBdSolInt}
\vec{\LB} = \LBp + Z \cH.
\end{align}
%

\section{Exact modification of the Boltzmann distribution at interfaces due to an electric field}
\label{ap:BoltzmannSources}
%
The previous section describes how to solve the linearized Boltzmann equation in some bulk region.  The total solution consists of the particular solution and a linear combination of the homogeneous solutions.  The electric field fixes the strength of the particular solution, while boundary conditions additionally determine the coefficients of expansion for the homogeneous solutions (given by $c_n$).  For bilayer systems, the scattering coefficients introduced in appendix \ref{ap:RTCoeff} provide the appropriate boundary conditions at the interface.  They relate the incoming and outgoing parts of the distribution functions.  However, the incoming and outgoing parts of the particular solution do not obey these boundary conditions.  Thus, one must construct the correct linear combination of homogeneous solutions (which form a complete set) to guarantee that the total distribution function satisfies interfacial boundary conditions.  The total solution changes if the electric field changes, in part because the electric field modifies the particular solution.  However, to continue satisfying the boundary conditions at the interface, the coefficients of expansion must change as well.  Thus, for bilayer systems, an external electric field modifies both the particular solution \emph{and} the coefficients of expansion.  

In appendix \ref{ap:SimplifiedForm} we derive the conductivity and torkivity tensors by approximating the non-equilibrium distribution function at the interface.  There we assumed that the particular solution sufficiently described the non-equilibrium distribution function that results from an external electric field.  By determining how the coefficients of expansion change in the presence of an external electric field, we obtain a more sophisticated ansatz of that distribution function.  Using the same procedure presented in appendix \ref{ap:SimplifiedForm}, but replacing the particular solution with this more sophisticated ansatz, one may obtain conductivity and torkivity tensors that better reproduce the physics of the Boltzmann equation.  We emphasize that this approach does not require one to completely solve the Boltzmann equation for the bilayer, but is far more computationally intensive than the approach outlined in appendix \ref{ap:SimplifiedForm}.

In the following we consider two regions separated by an interface, and extract the exact portion of the Boltzmann distribution modified by an external electric field.  For a given layer, $\LBps$ and $Z$ characterize the general Boltzmann distribution.  The electric field $E$ and the coefficients of expansion $c_n$ remain undetermined.  Here we require that the $\vec{k}$-space mesh of both regions contain $\Nt$ points. Thus, one may split any function defined on either Fermi surface into incoming and outgoing pieces, each of which contain $\Nt/2$ elements.

In general, one can model interfacial scattering in terms of an S-matrix, defined by
\begin{align}
\label{eq:SM1}
\begin{pmatrix}
\vec{\LB}^{\o}(0^-) \\
\vec{\LB}^{\o}(0^+)
\end{pmatrix}
=
S
\begin{pmatrix}
\vec{\LB}^{\i}(0^-) \\
\vec{\LB}^{\i}(0^+)
\end{pmatrix}
\end{align}
where $\vec{\LB}^{\i}(0^\pm)$ and $\vec{\LB}^{\o}(0^\pm)$ denote vectors with dimension $\Nth$, and describe the incoming and outgoing distribution functions on each side of the interface.  The $\Nt \times \Nt$ S-matrix
\begin{align}
S =
\begin{pmatrix}
S^{- -} 		&		 S^{- +}		 \\
S^{+ -}		&		 S^{+ +}
\end{pmatrix}.
\end{align}
is defined as follows
\begin{align}
[S^{- -}]_{ij,\alpha\beta} &= R^{-}_{ij,\alpha\beta}		 	\\
[S^{- +}]_{ij,\alpha\beta} &= T^{-+}_{ij,\alpha\beta}		\\
[S^{+ -}]_{ij,\alpha\beta} &= T^{+-}_{ij,\alpha\beta}		\\
[S^{+ +}]_{ij,\alpha\beta} &= R^{+}_{ij,\alpha\beta}.
\end{align}
Here $T^{\pm\mp}_{ij,\alpha\beta}$ and $R^{\pm}_{ij,\alpha\beta}$ give components of the S-matrix.  They equal the reflection and transmission matrices introduced in Appendix~\ref{ap:RTCoeff}.

Since the distribution function includes no quantum phase information, one cannot assume its continuity at the interface (i.e. $\vec{\LB}(0^-) \ne \vec{\LB}(0^+) $).  In order to obtain the solution of \eqref{eq:DBE2}, we must solve for the coefficients of expansion in each region such that the total solution satisfies the scattering matrix.  To accomplish this we write \eqref{eq:LBdSolInt} for both regions in terms of the incoming and outgoing parts:
\begin{align}
\begin{pmatrix}
\vec{\LB}^{\i}(0^-) \\
\vec{\LB}^{\i}(0^+)
\end{pmatrix}
&=
\begin{pmatrix}
\vec{\LB}^{\i}_{P}(0^-) \\
\vec{\LB}^{\i}_{P}(0^+)
\end{pmatrix}
\nonumber	\\[0.15cm]
&+
\begin{pmatrix}
Z^{\i}(0^-) 	& 	0 \\
0			&	Z^{\i}(0^+)
\end{pmatrix}
\begin{pmatrix}
\vec{c}(0^-) \\
\vec{c}(0^+)
\end{pmatrix},
\label{eq:fiexpansion}
\end{align}
\begin{align}
\begin{pmatrix}
\vec{\LB}^{\o}(0^-) \\
\vec{\LB}^{\o}(0^+)
\end{pmatrix}
&=
\begin{pmatrix}
\vec{\LB}^{\o}_{P}(0^-) \\
\vec{\LB}^{\o}_{P}(0^+)
\end{pmatrix}
\nonumber	\\[0.15cm]
&+
\begin{pmatrix}
Z^{\o}(0^-) 	& 	0 \\
0			&	Z^{\o}(0^+)
\end{pmatrix}
\begin{pmatrix}
\vec{c}(0^-) \\
\vec{c}(0^+)
\end{pmatrix}.
\label{eq:foexpansion}
\end{align}
The vectors $\vec{c}(0^\pm)$ contain the coefficients of expansion corresponding to the distribution functions at $z = 0^\pm$.  Notice that the same coefficients appear in both the incoming and outgoing equations.  In analogy to \eqrefs{ZDef}{cDef}, both $Z^{\i}(0^\pm)$ and $Z^{\o}(0^\pm)$ denote $\Nth \times \Nt$ matrices constructed from the column vectors $\vec{\EV}^{\i}(0^\pm)$ and $\vec{\EV}^{\o}(0^\pm)$ respectively.

Invoking the following convention for any vector $\vec{h}$ and matrix $H$
\begin{align}
\label{eq:VectorMatrixConvention}
\vec{h} = 
\begin{pmatrix}
\vec{h}(0^-) \\
\vec{h}(0^+)
\end{pmatrix}
\text{~~~~~~~~}
H = 
\begin{pmatrix}
H(0^-) 	& 	0 \\
0		&	H(0^+)
\end{pmatrix}
\end{align}
we may write Eqs.~(\ref{eq:SM1}), (\ref{eq:fiexpansion}), and (\ref{eq:foexpansion}) more compactly as
\begin{align}
\label{eq:SM2}
\LBout - S \LBin = \vec{0}
\end{align}
\begin{align}
\label{eq:fiexpansion2}
\vec{\LB}^\i
=
\LBpi
+
Z^\i
\vec{c}
\end{align}
\begin{align}
\label{eq:foexpansion2}
\vec{\LB}^{\o}
=
\LBpo
+
Z^\o
\vec{c}.
\end{align}
Note that $Z^\i$ and $Z^\o$ are $\Nt \times 2\Nt$ matrices.  Together, \eqrefss{eq:SM2}{eq:foexpansion2} provide us with a system of $\Nt$ equations to solve for $\vec{c}$.  However, $\vec{c}$ contains $2\Nt$ coefficients.  Without knowing the outer boundary conditions, one can only solve for half of the coefficients in terms of the other half.  We therefore separate $\vec{c}$ into the set of determined $\vec{c}_D$ and undetermined $\vec{c}_U$ coefficients, which gives:
\begin{align}
\vec{\LB}^{\i}
=
\LBpi
+ Z_D^\i \vec{c}_D
+ Z_U^\i \vec{c}_U
\end{align}
\begin{align}
\vec{\LB}^{\o}
=
\LBpo
+ Z_D^\o \vec{c}_D
+ Z_U^\o \vec{c}_U.
\end{align}
The matrices $Z_{U/D}^\io$ contain column vectors describing either the determined or undetermined solutions only.  As a result, they represent $\Nt \times \Nt$ matrices.   According to the convention established in \eqref{eq:VectorMatrixConvention}, both $\vec{c}_D$ and $\vec{c}_U$ are given by
\begin{align}
\vec{c}_D = \begin{pmatrix} \vec{c}_D(0^-) \\ \vec{c}_D(0^+) \end{pmatrix}
\text{~~~~~~~~}
\vec{c}_U = \begin{pmatrix} \vec{c}_U(0^-) \\ \vec{c}_U(0^+) \end{pmatrix},
\end{align}
and contain coefficients from each region.  Finally we define the vector
\begin{align}
\label{eq:bDef}
\vec{b}_A = \LB_A^\o - S \LB_A^\o 
\end{align}
which quantifies the extent to which the distribution $\LB^A$ satisfies the S-matrix.  For example, the $\vec{b}$ vector corresponding to the total distribution must vanish, since it satisfies the S-matrix.  One may equivalently write \eqref{eq:bDef} in terms of coefficients, i.e.
\begin{align}
\vec{b}_A = P_A \vec{c}_A
\end{align}
where
\begin{align}
P_A \equiv Z_A^\o - S Z_A^\i.
\end{align}

Using this notation we may rewrite \eqref{eq:SM1} as
\begin{align}
\LBout - S \LBin 
&= \vec{b}_P + \vec{b}_U + \vec{b}_D. \\
&= \vec{b}_P + P_U \vec{c}_U + P_D \vec{c}_D. \\
&= \vec{0}.
\end{align}
Solving for $\vec{c}_D$, we have:
\begin{align}
\label{eq:cD}
\vec{c}_D = T \vec{c}_U + \vec{c}_P
\end{align}
where
\begin{align}
T \equiv -[P_D]^{-1} P_U
\end{align}
\begin{align}
\vec{c}_P \equiv -[P_D]^{-1} \vec{b}_P.
\end{align}
\Eqref{eq:cD} implies the following: if one knows half of the coefficients, the remaining coefficients are related by the matrix $T$ (given that $P_D$ is invertible), in addition to a piece $\vec{c}_P$ caused solely by the electric field.  The coefficients contained within $\vec{c}_P$ give the desired modifications to the coefficients of expansion that are caused by an external electric field.

The portion of the incoming Boltzmann distribution caused by an external electric field then become
\begin{align}
\LB^\i_{\ep} 	&= \LB^\i_{P} + Z^\i_D \vec{c}_{P}.	\label{eq:LBSplitEField}
\end{align}
Recalling the convention set by \eqref{eq:VectorMatrixConvention}, the vector $\LB^\i_{\ep}$ includes distribution functions from both sides of the interface:
\begin{align}
\LB^\i_{\ep} = 
\begin{pmatrix}
\LB^\i_{\ep}(0^-) \\
\LB^\i_{\ep}(0^+)
\end{pmatrix}.
\end{align}
The remaining portion of the incoming Boltzmann distribution (independent of an external electric field) is given by
\begin{align}
\label{eq:LBSplitDetCoeff}
\LB^\i_\ip 	&= [ Z^\i_U + Z^\i_D T ] \vec{c}_U. 	
\end{align}
The external electric field $E$ and the undetermined coefficients $[\vec{c}_U]_n$ serve as input parameters to the full Boltzmann distribution; the remaining quantities in \eqrefs{eq:LBSplitEField}{eq:LBSplitDetCoeff} depend on material properties of the bulk regions and the interface.  In other words, $E$ and $[\vec{c}_U]_n$ now furnish the only degrees of freedom remaining in the interfacial Boltzmann distributions.  

We now discuss how to use this result to improve the conductivity and torkivity tensors.  We remind the reader that in order to derive those tensors, one must approximate the non-equilibrium distribution function $\LB^\i_{\alpha}(0^\pm, \kp)$ caused by an external electric field.  In appendix \ref{ap:SimplifiedForm} we approximate $\LB^\i_{\alpha}(0^\pm, \kp)$ using analytical expressions for the particular solutions, which were given by \eqrefs{eq:apIncDist1}{eq:apIncDist2}.  However, the vectors $\LB^\i_{\ep}(0^\pm)$ derived here are discrete representations (over momentum space) of the \emph{exact} distribution functions caused by an external electric field.  Thus, one could obtain $\LB^\i_{\alpha}(0^\pm, \kp)$ numerically by computing $\LB^\i_{\ep}$, rather than using the particular solutions alone.  This more sophisticated ansatz can be used in place of \eqrefs{eq:apIncDist1}{eq:apIncDist2} when computing the conductivity and torkivity tensors.

\bibliography{apssamp}


\end{document}